\documentclass[sigconf,screen]{acmart}
\AtBeginDocument{%
  }

\setcopyright{none}
\renewcommand\footnotetextcopyrightpermission[1]{}

\usepackage{booktabs}  
\usepackage{siunitx}   
\usepackage{graphicx}  
\usepackage{subcaption}
\usepackage{multirow}
\usepackage{url}
\usepackage{enumitem}
\begin{document}

\title{Agent4RE: A Self-Refining Multi-agent Framework for End-to-End Software Requirements Engineering and Benchmarking}

\author{Yongjian Tang}
\authornote{Corresponding author.}
\orcid{0009-0008-8670-6843}
\affiliation{%
  \institution{Siemens AG (RPD)}
  \country{Germany}
}
\affiliation{%
  \institution{Technical University of Munich}
  \city{Munich}
  \country{Germany}
}

\author{Linhan Li}
\orcid{0009-0005-9343-3992}
\affiliation{%
  \institution{Technical University of Munich}
  \city{Munich}
  \country{Germany}
}

\author{Thomas Runkler}
\orcid{0000-0002-5465-198X}
\affiliation{%
  \institution{Siemens AG}
  \country{Germany}
}
\affiliation{%
  \institution{Technical University of Munich}
  \city{Munich}
  \country{Germany}
}

\renewcommand{\shortauthors}{Tang, Li, Runkler}

\begin{abstract}
Existing LLM-based approaches for software Requirements Engineering (RE) typically rely on basic prompting strategies or rudimentary agent collaboration, under-utilizing the full potential of multi-agent systems. Meanwhile, available datasets focus on isolated subtasks, such as requirements extraction, classification, and completeness detection, leaving the absence of an end-to-end RE benchmark that spans from requirements elicitation to generation. We present Agent4RE -- a self-refining multi-agent RE system that orchestrates specialized agents and incorporates two iterative improvement loops. To support evaluation, we construct RE-E2E\footnote{We provide RE-E2E benchmark dataset, Agent4RE source code, and supplementary materials at \url{https://github.com/markustyj/Agent4RE_ASE-2026}.} -- a real-world dataset built from human-written requirement specifications, enabling end-to-end assessment of RE workflows. Building on this foundation, we further propose two enhanced Agent4RE versions that incorporate either autonomous self-refinement or structured human feedback, and analyze their strengths and limitations across different scenarios. Evaluation on 8 Large Language Models (LLMs) demonstrates that all three Agent4RE variants consistently outperform a domain-context-augmented prompting baseline by average 8\% in text-based metrics. The two enhanced variants achieve the highest LLM-as-a-judge and human ratings, surpassing two RE baselines by approximately 0.8 points on a four-point scale. This consistent performance establishes Agent4RE as a practical end-to-end RE solution for industrial environments.
\end{abstract}


  


\maketitle

\section{Introduction}
Requirements Engineering (RE) serves as a preliminary stage in software development, establishing a shared understanding of stakeholders' needs, goals, and constraints while providing the foundation for subsequent software design, implementation, and verification \cite{laplante2022requirements}. Traditionally, this process has required substantial effort from multiple roles -- developers, business analysts, domain experts, and project managers -- who must iteratively perform elicitation, analysis, specification, and validation together to produce complete, consistent, and clear requirement documentation \cite{nuseibeh2000requirements}. Given this labor-intensive nature and the fact that defects originating from this phase are significantly more expensive to fix in later stages, improvements in RE efficiency and quality can yield substantial benefits across the entire software development lifecycle. 

Recently, advancements in Large Language Models (LLMs) have opened new avenues for requirements engineering \cite{two-way_roadmap}. With superior capabilities in domain knowledge inference, natural language understanding, and ambiguity detection, LLMs are well suited to accelerate and assist diverse RE sub-tasks, such as extracting requirements from unstructured inputs, identifying inconsistencies in RE specifications, and generating documentation based on stakeholder instructions \cite{zadenoori2025large}. However, existing studies predominately focus on such isolated RE sub-tasks, achieving success only on individual activities, 
yet offering limited support for the complex, interconnected requirements engineering processes in real-world scenarios. While recent efforts have explored LLM-based interactive systems and agents to aggregate these isolated capabilities \cite{tang2026agent4csmultiagentcodesummarization}, no systematic end-to-end solution has been proposed and evaluated for the full RE lifecycle. Similarly, this fragmentation extends to data resources: accessible datasets are limited to separate RE tasks and no comprehensive open-source benchmark has been released to assess the entire requirements engineering process.

\begin{figure*}[h!]
  \centering
  \includegraphics[width=0.91\linewidth]{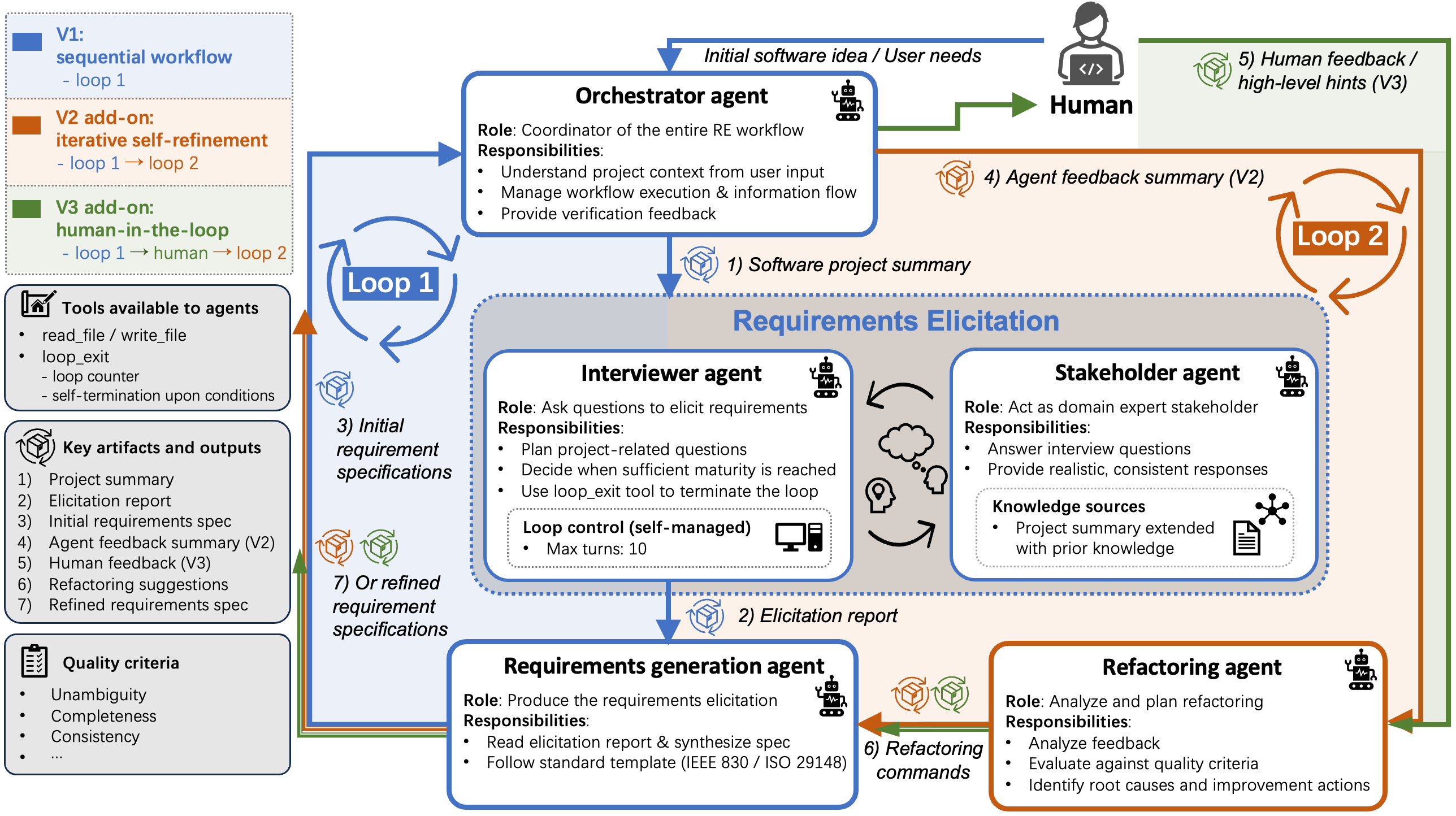}
  \caption{Agent4RE orchestrates specialized agents across two nested feedback loops: an inner generation loop that iteratively elicits requirements and drafts the initial specification, and an outer refactoring loop that continuously optimizes the specification through self-refinement. Agent4RE supports three operational modes: sequential workflow without feedback (blue), iterative self-refinement (orange add-on), and agentic workflow with human-in-the-loop (green add-on).
  Given a brief software project description from users, Agent4RE produces complete requirement specifications autonomously.
  }
  \label{fig:agent_overview}
  \Description{Agent4RE simulates the real-world requirements engineering process with five specialized agents: an orchestrator agent that understands the project scope by
analyzing the user message, an interviewer agent and stake-
holder agent that simulate requirements elicitation with self-
controlled elicitation loops, a generation agent that finalizes the requirement specifications, and a refactoring agent that iteratively optimizes the specification. It is the earliest attempt to solve end-to-end requirements engineering problem through iterative multi-agent workflows.}
\end{figure*}

To address this gap, we propose Agent4RE -- a self-refining multi-agent system for requirements engineering, which orchestrates specialized agents to simulate stakeholder interactions and user behaviors, enabling iterative elicitation, generation, and traceable refinement of requirements, as illustrated in Figure \ref{fig:agent_overview}. Combined with RE-E2E -- an end-to-end RE dataset, our framework mitigates the bottleneck of current LLM-based RE research, bridging isolated task-specific approaches with complete RE workflows.
The main contribution of this work can be summarized as follows: 
\begin{itemize}    
    \item [(i)]  
    We propose Agent4RE -- an autonomous multi-agent system for requirements engineering, which features two iterative improvement loops for self-refinement. 
    
    \item [(ii)] 
    We construct RE-E2E -- an RE benchmark built from full-scale human-written requirement specifications. To the best of our knowledge, this represents the first end-to-end dataset designed for complete requirements engineering processes, spanning from initial project descriptions to finalized requirement specifications.

    \item [(iii)] 
    We analyze the performance of Agent4RE by integrating eight frontier LLMs covering different performance tiers, studying its effectiveness and robustness across diverse model capabilities and computational budgets. Furthermore, we showcase the self-refinement features of all language models by analyzing their iterative improvement loops.

    \item [(iv)] 
    We validate our approach by comparing the generated requirement specifications with the ground-truth in RE-E2E. Evaluation results reveal that all Agent4RE variants consistently outperform a context-augmented prompting baseline by average 8\% in text-based metrics. Furthermore, the two enhanced variants achieve the highest LLM-as-a-judge and human ratings, surpassing two baselines by approximately 0.8 points on a four-point scale.
    
\end{itemize}

\section{Related Work}

\subsection{Large Language Models for RE-related Tasks}

Extensive language model-based approaches have been proposed for diverse stages throughout the requirements engineering cycle. Building on the early transformer architecture \cite{vaswani2017attention}, researchers fine-tuned BERT to detect anaphoric ambiguity in requirements and evaluated it against traditional supervised entity recognition techniques \cite{ezzini2022automated}.  
A similar approach was introduced to identify intra-domain ambiguities in the requirements of interdisciplinary projects \cite{identification_ambiguity}. Given BERT's encoder-based architecture, parallel studies trained specialized BERT variants for requirements classification, achieving improved accuracy over machine learning techniques like decision trees and support vector machines \cite{norbert,prcbert}.  
As language models have scaled up, strategic few-shot prompting has emerged as a more streamlined solution to requirements classification, delivering on-par performance with fine-tuned models while eliminating the computational overhead of training \cite{el2023ai,tang2024fsponer,tang2025few}.

The superior reasoning capabilities of these large-scale models enable effective requirements analysis while their generative nature aligns with the needs for requirements elicitation and specification generation. Given this potential, a preliminary evaluation \cite{arora2024advancing} explored the role of LLMs across different RE stages and suggested tailored prompt patterns for elicitation, analysis, specification, and validation tasks. Building on this foundation, LLMREI \cite{korn2025llmrei} employs LLMs as automated interviewers within a chatbot to simulate the requirements elicitation process with human stakeholders, designed to minimize human intervention and mitigate common interview-related errors. 
SpecGen \cite{ma2025specgen} leveraged a two-phase prompting approach to generate precise behavioral comments for program-level specifications, while ReqGPT \cite{Schiller_Haddad_Seibel_2025}, a 7B-parameter domain-specific LLM fine-tuned on over one hundred curated requirement lists, was developed to improve specification quality. 
Despite these breakthroughs in specific RE aspects, existing methods still fall short in delivering complete requirement specifications for production environments, indicating the need for more robust and comprehensive approaches.

\subsection{LLM-based Agentic Systems for Automated RE Process}
The collaborative workflow of multi-agent systems offers a promising paradigm to tackle more complex end-to-end RE challenges \cite{tang2026llm}. As an early exploration \cite{early_mas4re}, researchers implemented an LLM-based agentic system with GPT-3.5 and GPT-4 to automate user story elicitation, requirements generation, and prioritization for agile software projects. Extending this prior work, a subsequent study proposed a role-based multi-agent system that emulates stakeholder roles to collaboratively generate and prioritize requirements from project descriptions \cite{automated_requirements_analysis}. 
Another modular architecture ReqInOne breaks down requirements generation into three sub-tasks: summary, extraction, and classification \cite{reqinone}, using tailored prompt templates to improve the quality and consistency of LLM outputs.
In comparison, Elicitron has focused on more specific RE activities \cite{ataei2024elicitron}. The system simulates elicitation interviews to explore product experiences and identify stakeholders' needs, thus enabling the discovery of unforeseen use cases through structured LLM interactions. 
Moving toward end-to-end automation, iReDev \cite{jin2025iredev}, previously also called MARE \cite{jin2024mare}, represents a knowledge-driven multi-agent framework designed to address the full requirements engineering process. It incorporates human expert knowledge into six agents via chain-of-thought prompting and structured reasoning, enabling them to manage an event-driven artifact pool collaboratively. 
Nevertheless, due to the absence of resources for end-to-end RE assessment, the evaluation focused primarily on intermediate steps such as requirements modeling and entity extraction. The final specifications were evaluated only through subjective metrics such as completeness, correctness, and consistency.

\subsection{Baseline Selection}
Given our focus on end-to-end requirements engineering, approaches that specifically address intermediate RE steps or sub-tasks, such as requirements elicitation and prioritization, are poorly suited and insufficient for our purposes, as illustrated in Table \ref{tab:literature_comparison}. 
Therefore, we select the only two existing baselines that address the complete RE lifecycle: a prompting-based approach with tailored prompt patterns for different RE stages \cite{arora2024advancing}, and iReDev, a multi-agent collaboration framework for RE \cite{jin2025iredev}. 

\begin{table}[ht]
\centering
\caption{Application domains of selected BERT variants, LLM-based approaches, and agentic solutions for RE.}
\label{tab:literature_comparison}

\resizebox{1\linewidth}{!}{
\setlength{\tabcolsep}{10pt}
\begin{tabular}{l| c c c c}
\toprule
\textbf{Approaches} & \multicolumn{4}{c}{\textbf{Covered RE stages} } \\
 & Elicitation & Analysis & Specification & Validation \\
\midrule

BERT ambiguity \cite{ezzini2022automated} &  & \checkmark &  & \\
NoRBERT \cite{norbert} &  & \checkmark &  & \\
PRCBERT \cite{prcbert}  &  & \checkmark &  & \\
\midrule   
Over-prompting \cite{tang2025few} &  & \checkmark &  & \\
Prompt patterns \cite{arora2024advancing} & \checkmark & \checkmark &  \checkmark & \checkmark\\
 LLMREI \cite{korn2025llmrei} & \checkmark &  &  & \\
 SpecGen \cite{ma2025specgen} &  & \checkmark & \checkmark & \\
 ReqGPT \cite{Schiller_Haddad_Seibel_2025} &  & \checkmark & \checkmark & \\
 \midrule
 Early agents \cite{early_mas4re} & \checkmark & \checkmark &  & \\
 Role-based agents \cite{automated_requirements_analysis}& \checkmark & \checkmark & \checkmark & \\
 Elicitron \cite{ataei2024elicitron} & \checkmark &  &  & \\
 ReqInOne \cite{reqinone} &   & \checkmark &  \checkmark &  \\
 iReDev \cite{jin2025iredev} & \checkmark & \checkmark &  \checkmark & \checkmark\\
 \textbf{Agent4RE} & \checkmark & \checkmark &  \checkmark & \checkmark\\
\bottomrule

\end{tabular}
}

\end{table}
\section{Agent4RE}
Our multi-agent system offers three different versions, each with distinct advantages: 1) a sequential workflow without feedback for streamlined RE processes; 2) a workflow with self-refinement that enables the system to iteratively improve the outputs; 3) a workflow with human feedback that incorporates domain experts to guide the refinement process.

\subsection{Sequential Workflow without Feedback}
\begin{figure}[ht]
  \centering
  \includegraphics[width=1\linewidth]{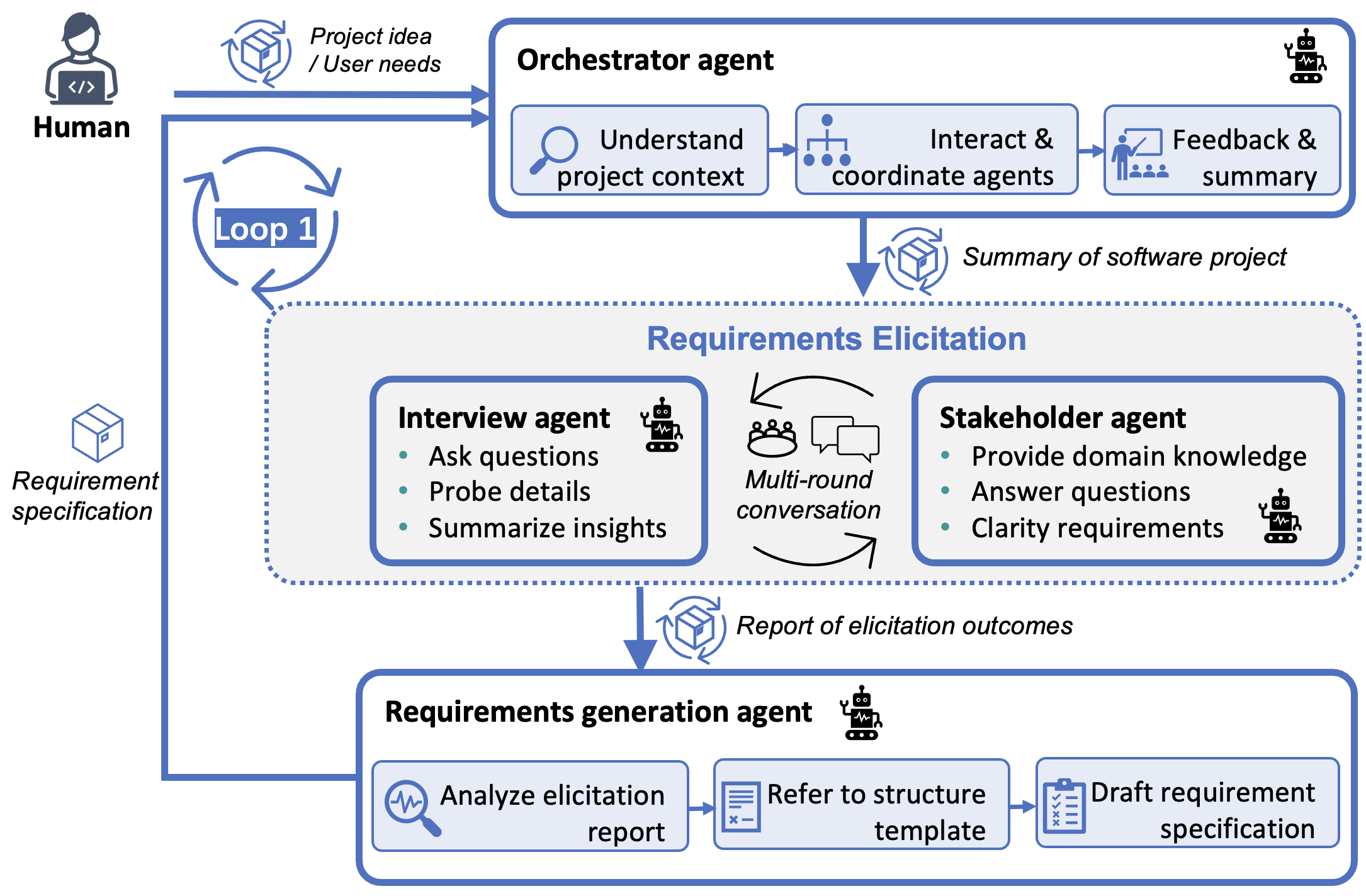}
  \caption{The sequential workflow of Agent4RE consists of an orchestrator agent that understands the project scope by analyzing the user message, an interviewer agent and stakeholder agent that simulate requirements elicitation with self-controlled elicitation loops, and a generation agent that finalizes the requirement specifications.}
  \Description{This is the first sequential loop to generate the initial draft of a requirement specification.}
  \label{fig:agent_overview_loop1}
\end{figure}

Figure \ref{fig:agent_overview_loop1} illustrates the first version: a sequential workflow of Agent4RE. The multi-agent system starts with an orchestrator agent $A_O$ \linebreak\mbox{responsible} for communicating and interacting with human users. Once receiving a brief project idea $I$, the orchestrator analyzes and understands the project context using its pre-trained knowledge. Project summary $S$ will be generated to guide subsequent procedures: $S = A_O(I)$.

The iterative requirements elicitation process represents the core step of this workflow and is self-controlled by an interviewer agent $A_I$ and a stakeholder agent $A_S$. Given a message from the orchestrator agent, the interviewer agent initiates the conversation by posing a list of project-related questions $q$. Meanwhile, the stakeholder agent familiarizes itself with the project domain by reading the orchestrator-generated project summary and combing its own knowledge base to answer interviewers' questions. 
\begin{table}[t]
\centering
\caption{Core Notation and Functional Mappings.}
\label{tab:core_notation}
\resizebox{1\linewidth}{!}{  
\setlength{\tabcolsep}{3pt}  
\begin{tabular}{ll}
\toprule
\textbf{Notation} & \textbf{Description / Functional Definition} \\
\midrule
$I$ & The idea of raw project 
\\
$S$ & The summary of software project generated by the orchestrator \\ 
$q_t, a_t$ & Elicitation questions and answers exchanged at turn $t$ \\
$R$ & The report summarizing the elicitation history \\
$T_{re}$ & The template for generated requirement specifications \\
$D_k$ & The document of requirement specification at refinement iteration $k$ \\
$F_a$, $F_h$ & The set of agent and human feedback in bullet points \\
\midrule
$A_O(I)$ & \textbf{Orchestrator Agent:} Analyzes context and validates deliverables \\
$A_I(q_{t-1},a_t, S)$ & \textbf{Interviewer Agent:} Generates exploratory elicitation questions \\
$A_S(q_t, S)$ & \textbf{Stakeholder Agent:} Simulates answers using its domain knowledge  \\
$A_G(R, T_{re})$ & \textbf{Generator Agent:} Synthesizes or refines the specification document \\
$A_R(D, F_a, F_h)$ & \textbf{Refactor Agent:} Formulates targeted quality-improvement commands \\
\bottomrule
\end{tabular}
}
\end{table} 
This simulated interview mirrors real-world elicitation scenarios and continues for multiple rounds until the discussion outcomes reach sufficient maturity. Using the core notation and functional mappings listed in Table \ref{tab:core_notation}, each iteration follows the process: 
\begin{equation}
    q_t = A_I(q_{t-1}, a_{t-1}, S);   \quad a_t = A_S(q_t, S).
\end{equation}

To ensure self-management, we equip the interviewer agent with a loop exit tool that enables autonomous termination of the elicitation process. Notably, the tool incorporates a loop counter, with a pre-set maximum of 10 iterations to avoid redundant repetition in practice. In the end, the interviewer summarizes the entire conversation into an elicitation report $R$:
\begin{equation}
\resizebox{0.44\textwidth}{!}{
$R = \text{Summarize}\big(\{(q_1, a_1), (q_2, a_2), \dots, (q_t, a_t)\}\big), \; t \le 10.$
}
\end{equation}

With the detailed report, the requirements generation agent $A_G$ produces the initial specification document: $D_0 = A_G(R, T_{re})$.
To ensure structural compliance, the required specification templates $T_{re}$ \cite{template_1998,template_2018} are integrated into LLMs' system message as a reference, steering the agent to strictly follow the structure. Due to the document length, the generation time typically ranges from a few seconds to over one minute, depending on the model size and inference configuration. Once finished, the generated specification is returned to the orchestrator for validation, which verifies the document before handing over to human users. In case of any deficiencies, it leverages its supervisory control over other agents to perform corrective action. Further investigation of this self-controlled iterative elicitation loop is provided in Section \ref{sec:loop_experiments}.

\subsection{Agentic Workflow with Self-refinement}
As illustrated in Figure \ref{fig:agent_overview}, an additional refinement loop (in orange) is introduced to further enhance the quality of the generated specifications. A refactoring agent $A_R$ serves as the central component in this process, analyzing the verification results from the initial loop (in blue) and applying pre-defined requirements quality criteria such as unambiguity, completeness, and consistency to formulate improvement suggestions. 
These generated refactoring suggestions are then forwarded to the generation agent, which refines the original specification accordingly. This agent-based self-refinement loop can also be executed multiple times until the specification meets the demanded quality standards, which can be formally represented as: $A_R\big(D_k, F_a\big)$, where $D_k$ represents the generated specification document in the $k$-th round and $F_a$ the set of feedback from agents.

\subsection{Agentic Workflow with Human Feedback}
While automated agentic workflows streamline the RE process, including human-in-the-loop at critical decision points can further enhance the quality and reliability of the generated specifications \cite{natarajan2025human}. Therefore, in the third version of Agent4RE, we incorporate human feedback $F_h$ after the initial sequential loop and prior to the refactoring loop. 
In this case, human developers provide domain-specific insights and expertise alongside the verification comments from the orchestrator agent to guide the refactoring agent. In our experiments, all human participants are experienced software engineers and they have access to the ground-truth specifications from the annotated dataset. 
However, to simulate realistic scenarios, human feedback is limited to concise bullet points or high-level hints regarding specification structure rather than detailed corrections. 

The progression of the document from state $D_k$ to $D_{k+1}$ across the three distinct versions of Agent4RE can be formulated as:
\begin{equation}
    D_{k+1} = A_G\Big(D_k, \; A_R\big(D_k, F_a, F_h\big)\Big).
\end{equation}

\subsection{Implementation and Configuration}
\begin{figure}[h]
  \centering
  \includegraphics[width=0.88\linewidth]{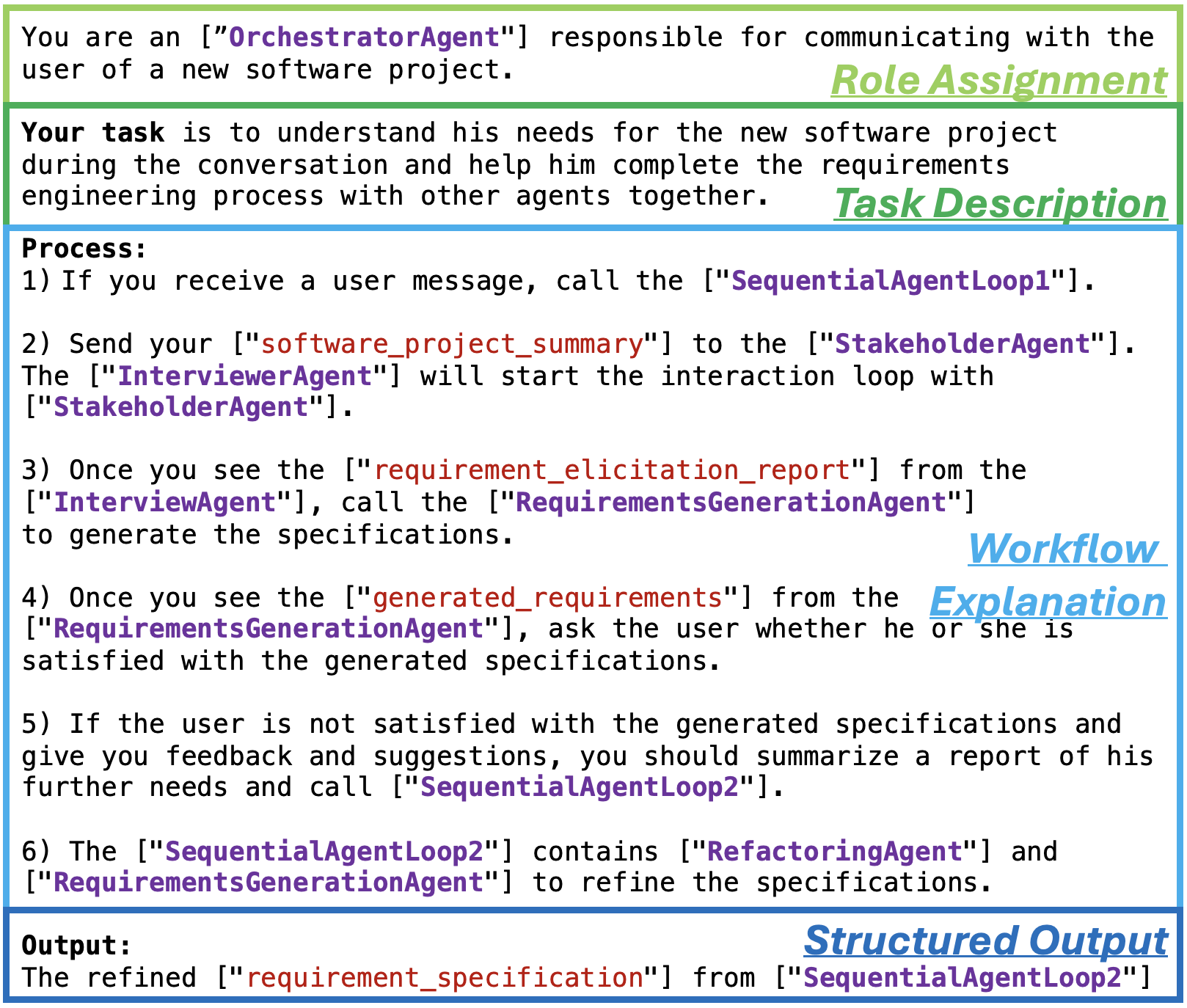}
  \caption{The system prompt for the orchestrator agent used in self-refinement with human feedback.}
  \Description{The orchestrator prompt is used to explain the entire RE process to orchestrator agent. This determines the smooth execution of workflow and coordination among agents.}
  \label{fig:orchetrator_prompt}
\end{figure}

We implement Agent4RE using the Google AI Development Kit (ADK), which provides native support for agent coordination and workflow management \cite{adk}. Each agent in our system is instantiated as an independent agent with specialized prompts and tool configurations. For instance, the orchestrator agent -- responsible for managing and coordinating all specialized agents -- operates using the system prompt demonstrated in Figure \ref{fig:orchetrator_prompt}. Due to space constraints, the complete prompt templates for the remaining agents are provided in the supplementary materials.


Unlike previous agentic solutions that apply uniform LLM settings, we adopt task-specific temperature configurations for all agents, as detailed in Table \ref{tab:agent_temperature}. Temperature controls the randomness of the model's output, with lower values producing more deterministic responses and higher values encouraging creativity. This differentiated approach allows Agent4RE to maximize each agent's potential for their task.

\begin{table}[ht]
\centering
\caption{Configuration of agent temperature.}
\label{tab:agent_temperature}
\resizebox{0.87\linewidth}{!}{
\setlength{\tabcolsep}{8pt}
\begin{tabular}{ l l l }
\toprule
\multicolumn{2}{l}{Agents\&Temperature} & Consideration\\
\midrule
Orchestrator & 0.5 & Balanced temperature for coordination tasks\\
Interviewer & 1 & Higher temperature for exploratory elicitation\\
Stakeholder & 1 & Higher temperature for exploratory elicitation \\
Generator & 0.8 & Higher temperature for creative generation \\
Refactor & 0.2 & Lower temperature for consistent suggestion\\
\bottomrule
\end{tabular}
}
\end{table}

\section{Experiments}
\subsection{RE-E2E Dataset}

\begin{table*}[ht]  
\centering  
\caption{Statistics of the requirement specification dataset.}  
\label{tab:dataset_statistics}  
\resizebox{0.8\linewidth}{!}{  
\setlength{\tabcolsep}{15pt}  
\begin{tabular}{l S[table-format=5.2]}  
    \toprule  
    \multicolumn{2}{c}{\textbf{IEEE 830-1998}} \\ Statistics & {Value} \\  
    \midrule  
    Total dataset tokens & 107887 \\  
    Number of collected files & 30 \\  
    Average letters per file & 18462.43 \\  
    Average tokens per file & 3596.23 \\  
    Average sections per file & 37.97 \\  
    Average pages per file & 20.53 \\  
    Rate of non-empty sections (\%) & 84.23 \\  
    \bottomrule  
\end{tabular}  
\hspace{2.5cm} 
\begin{tabular}{l S[table-format=5.2]}  
    \toprule  
    \multicolumn{2}{c}{\textbf{IEEE 29148-2018}} \\  
    Statistics & {Value} \\  
    \midrule  
    Total dataset tokens & 72157 \\  
    Number of collected files & 20 \\  
    Average letters per file & 17673.90 \\  
    Average tokens per file & 3607.85 \\  
    Average sections per file & 45.15 \\  
    Average pages per file & 22.85 \\  
    Rate of non-empty sections (\%) & 79.12 \\  
    \bottomrule  
\end{tabular}}  
\end{table*}

Existing data resources for RE predominantly focus on individual stages, suffering from several limitations: incomplete coverage, poor community visibility, and limited support for end-to-end evaluation \cite{motger2025}. For instance, PROMISE \cite{promise}, a requirements classification benchmark extensively used in prior work, contains only isolated requirement items. Similarly, PURE \cite{pure} provides requirement specification documents for various software systems, but lacks a unified structural template for standardized processing and systematic evaluation. And a more recent RE dataset \cite{data_domain_models} provides a collection of user stories pairing with the corresponding domain models, yet it still excludes complete requirement specifications. As LLM capabilities and agentic workflows advance continuously, the ambition within the research community and industry has grown beyond addressing isolated RE tasks toward autonomous management of the entire RE lifecycle. Addressing data scarcity has become an urgent priority to enable the exploration of end-to-end agentic solutions.

To address this research gap, we curate RE-E2E, a dataset used to benchmark the performance of different frameworks for end-to-end software requirement engineering tasks. We selectively collect 50 complete requirement specifications in PDF format from real-world software repositories \cite{srs}. Detailed statistics of the files are presented in Table \ref{tab:dataset_statistics}. All files originate from experienced software engineers and cover various software domains, such as software control, e-commerce, and security verification systems, and translation software.

Based on the original specification template used by developers, we divide the collected PDF files into two sets: one adhering to IEEE 830 \cite{template_1998}, which defines the structure for traditional software requirements specifications, and another following IEEE 29148 \cite{template_2018}, an updated standard that incorporates systems and software engineering requirements with more modern practices.
Specifically, we use regular expressions \cite{friedl2006mastering} to clean noise such as unexpected newline characters and consecutive spaces, obtaining normalized plain text. 
However, in some edge cases, the original section names of human-written specifications do not strictly follow the IEEE format. To create a dataset with unified structure, we match these sections with the official IEEE section names based on their semantic embeddings. To ensure dataset reproducibility, we use the open-source Qwen3-embedding-8B to match RE-sections. If the cosine similarity between two sections falls below 0.7, human annotators must intervene. 
They review each section independently and reach consensus on whether to remove or retain it as empty. The cross-annotator agreement is 96.7\%, meaning that all annotators make the same binary decision in most cases. For the remaining 3.3\%, annotators follow majority vote. In Table \ref{tab:dataset_statistics}, we record the section empty rates. 

Based on the finalized specifications, human annotators manually write the initial project descriptions, which are then syntactically checked by LLMs to ensure grammatical correctness and formatting compliance. Notably, these annotators also serve as human evaluators in experiments. Their prior experience with the annotation process ensures their familiarity with the content. Further details about their information are provided in Section \ref{metric_section_human}.
The resulting project descriptions serve as inputs for Agent4RE. The processed descriptions and software requirement specifications are provided in the supplementary materials.

\subsection{Selected Language Models}
Our experiments involve eight LLMs spanning different performance tiers to investigate the full potential of our agentic framework. We deploy four frontier models via APIs, including GPT-5 \cite{2025gpt5}, GPT-4.1 \cite{2024gpt41}, GPT-4o \cite{hurst2024gpt4o}, and Gemini-3-flash \cite{comanici2025gemini}, which represent the forefront of LLM capabilities. To complement these proprietary models, we incorporate four open-source alternatives that ensure reproducibility and broader applicability: LLaMA3.1-8B from Meta AI \cite{dubey2024llama}, Qwen3-8B from Alibaba \cite{yang2025qwen3technicalreport}, Ministral3-8B from Mistral \cite{mistral3}, and Gemma4-4B from Google \cite{gemma4}. 

\subsection{Metrics}
\label{metric_section_human}
To compare the generated requirement specifications with the ground-truth in RE-E2E, we employ an extensive set of evaluation approaches, including text-based metrics, semantic-based metrics, LLM-as-a-judge, and human evaluators.

\begin{figure}[h]
  \raggedleft
  \includegraphics[width=0.89\linewidth]{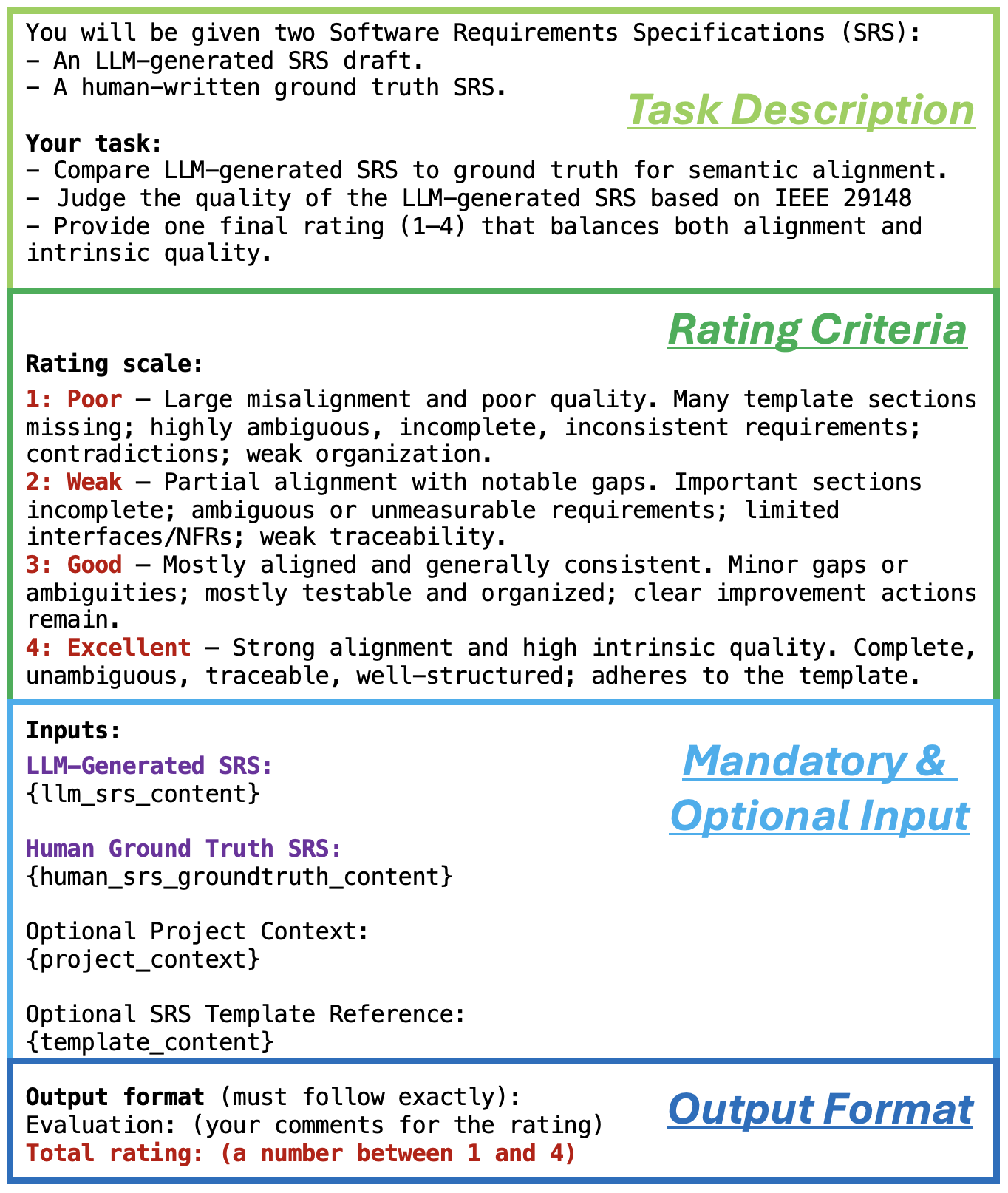}
  \caption{The prompt used for LLM-as-a-judge spans four rating scales, each with detailed criteria to assess requirements on ambiguity, completeness, and other dimensions.}
  \Description{This is the prompt used for judging the quality of generated requirement specifications, covering four different rating scales, each with detailed assessment criteria.}
  \label{fig:prompt_llm_as_a_judge}
\end{figure}

\begin{itemize}
    \item [1)] Regarding text-based metrics, we apply \textbf{METEOR} \cite{banerjee2005meteor}, \textbf{BLEU-1} \cite{papineni2002bleu}, and \textbf{ROUGE-L} \cite{lin2004rouge}. 
    METEOR evaluates text similarity through a harmonic mean of precision and recall, incorporating stemming, synonymy, and word order penalties. 
    BLEU-1 
    measures the proportion of individual words (unigrams) in the generated text that match the reference text, while ROUGE-L evaluates the longest common subsequence between the candidate and reference texts.

    \item [(2)] For semantic-based metrics, we apply \textbf{BERTScore} \cite{zhang2019bertscore} and \textbf{SentenceBert} \cite{reimers2019sentence}. 
    BERTScore computes token-level semantic similarity by comparing BERT-based word embeddings between the candidate and reference texts, while SentenceBert measures sentence-level similarity by encoding texts into dense vector representations and calculating their cosine similarity.
    
    \item [(3)] \textbf{LLM-as-a-judge}
    leverages prompt-based assessment to grade the generated requirement specifications. 
    Figure \ref{fig:prompt_llm_as_a_judge} presents the prompt template used for this approach, with grades assigned on a scale from 1 to 4. 
    To assess RE quality in practice, developers typically require a language checker for ambiguity, an app-lifecycle tool for completeness, and a formal/model checker for logical consistency, leading to increased toolchain complexity. 
    Since these quality criteria are fundamentally grounded in written text and requirements, LLMs are natural requirements evaluators given their expertise in language processing. 
    However, this suitability does not apply to all model scales. While large-scale models demonstrate promising capabilities in judging text quality, compact models often struggle~\cite{gu2024survey} -- even the best one misjudges. Therefore, we deploy this evaluation strategy exclusively to four top-tier models (GPT-5, GPT-4.1, GPT-4o, and Gemini-3-flash). 
   
    \item [(4)] We further include \textbf{human} evaluation for comparison. To align with LLM-as-a-judge, three software developers from industry assess summaries independently on a scale from 1 to 4, with final scores averaged for fairness. We select human evaluators based on their software development experience and RE domain expertise. Finally, we choose three senior software engineers, each with two to three years of development experience in industry. In addition to their own judgment criteria, we also provide them with the prompt and grading schema used in LLM-as-a-judge to ensure consistent evaluation.

\end{itemize}

\subsection{RE Specification Generation}
Since iterative human feedback has no theoretical upper limit on performance, we perform one optimization loop for all Agent4RE variants to ensure fair comparison. 
Table \ref{tab:specification_big_table} demonstrates the evaluation results for end-to-end specification generation in the next page. Agent4RE significantly outperforms the two baselines in most considered metrics, with each variant exhibiting superior performance in different evaluation dimensions. For traditional text- and semantic-based metrics, incorporating high-level feedback from human developers yields the best results, achieving 0.251 in METEOR, 0.273 in Rouge-L, 0.063 in BLEU-1, 0.838 in BERT F1 score, and 0.392 in Sentence-BERT similarity. As for LLM-as-a-judge, pure self-refinement with GPT-5 slightly outperforms the human-in-the-loop variant, with both GPT-5 and GPT-4.1 models securing perfect 4.0 ratings when evaluated by GPT-4.1. However, upon human evaluation, self-refinement with human feedback still performs best, scoring the highest 3.29 out of 4. In comparison, the two baselines fall short, reaching 2.22 and 2.87 respectively in human evaluation.  

If we zoom in model performance for detailed analysis, three GPT models lead the performance in most metrics, especially when deployed within the sequential workflow without feedback or self-refinement with human feedback. 
Among them, the latest GPT-5 model demonstrates exceptional performance in LLM-as-a-judge and human evaluation, achieving near-perfect scores from other LLM evaluators when integrated into the Agent4RE framework. However, an unexpected finding emerges with pure self-refinement: iterative improvement loops lead to degraded performance on semantic-based metrics, as evidenced by Sentence BERT. This phenomenon may stem from LLMs' deliberate attempts to modify content, and such purposeful revisions can result in substantial semantic drift. We provide a case study in the supplementary materials, including an example of software project description, human feedback, and the generated requirement specifications. Due to the length of the generated content, a separate JSON file recording the output of each agent throughout the complete agentic workflow is included in the uploaded repository.

Detailed resource consumption for each model is reported in Table \ref{tab:cost_and_time}.
\begin{table*}[ht]  
\centering  

\caption{Comparison of generated requirement specifications with ground-truth data, evaluated using text-based, semantic-based, and LLM-based metrics, as well as human evaluations.}  
\label{tab:specification_big_table}

\resizebox{1\linewidth}{!}{
\setlength{\tabcolsep}{12pt}

\begin{tabular}{c c c c c c c c c c c c}  
\toprule
& & \multicolumn{3}{c}{\textbf{Text Based}} & \multicolumn{2}{c}{\textbf{Semantic Based}}  
& \multicolumn{4}{c}{\textbf{LLM-as-a-judge} } & \textbf{Human}\\  

\cmidrule(lr){3-5} \cmidrule(lr){6-7} \cmidrule(lr){8-11} \cmidrule(lr){12-12} 

\textbf{Baselines \& Agent4RE} & \textbf{Models} & METEOR & Rouge-L & BLEU-1 & BERTScore & Sentence- & GPT-4o & GPT-4.1 & GPT-5 & Gemini- & \\   

&  &  &  &  & (F1) & BERT &  &  &  & 3-flash & \\   
\midrule

 & GPT-5 & \textbf{0.183} & 0.102 & 0.014 & 0.810 & 0.378 & \textbf{3.12} & \textbf{3.28} & \textbf{2.24} & \textbf{3.20} & 2.13 \\  
 & GPT-4.1 & 0.145 & 0.117 & \textbf{0.015} & 0.806 & 0.350 & 3.04 & 3.10 & 1.76 & 3.02 & \textbf{2.22} \\ 
  & GPT-4o & 0.118 & 0.124 & 0.010 & 0.810 & \textbf{0.412} & 2.68 & 2.34 & 1.48 & 2.52 & 1.81 \\ 
 \textbf{Baseline} & Gemini-3-flash & 0.122 & 0.104 & 0.011 & 0.811 & 0.364 & 2.74 & 2.46 & 1.66 & 3.24 & 1.92 \\ 
 prompt patterns & LLaMA-3.1-8B & 0.133 & 0.131 & 0.012 & \textbf{0.817} & 0.363 & 2.56 & 2.34 & 1.48 & 1.96 & 1.52 \\ 
 & Qwen3-8B & 0.117 & \textbf{0.132} & 0.008 & 0.810 & 0.344 & 2.76 & 2.58 & 1.76 & 2.42 & 1.78 \\ 
 & Ministral3-8B & 0.110 & 0.115 & 0.005 & 0.788 & 0.332 & 3.02 & 2.96 & 1.72 & 2.92 & 1.82 \\ 
 & Gemma4-4B & 0.047 & 0.066 & 0.002 & 0.802 & 0.212 & 2.48 & 1.76 & 1.20 & 2.04 & 1.31 \\

\cline{2-12}
\addlinespace[4pt]
& GPT-5 & \textbf{0.197} & 0.102 & 0.012 & 0.816 & 0.374 & \textbf{3.84} & 3.88 & \textbf{2.62} & \textbf{3.76} & \textbf{2.87} \\   
& GPT-4.1 & 0.175 & 0.132 & 0.016 & 0.818 & 0.402 & 3.58 & \textbf{3.92} & 2.44 & 3.70 & 2.80 \\  
 & GPT-4o & 0.160 & \textbf{0.162} & \textbf{0.018} & \textbf{0.822} & \textbf{0.418} & 3.20 & 3.34 & 2.16 & 3.14 & 2.70 \\ 
\textbf{Baseline} & Gemini-3-flash & 0.158 & 0.136 & 0.015 & 0.815 & 0.376 & 3.36 & 3.54 & 2.22 & 3.72 & 2.72 \\  
iReDev (MARE) & LLaMA-3.1-8B & 0.102 & 0.128 & 0.008 & 0.821 & 0.342 & 2.36 & 2.40 & 1.60 & 2.52 & 2.02 \\  
& Qwen3-8B & 0.107 & 0.134 & 0.008 & 0.812 & 0.330 & 3.12 & 3.50 & 2.00 & 3.36 & 2.20 \\  
& Ministral3-8B & 0.095 & 0.118 & 0.005 & 0.805 & 0.398 & 3.26 & 3.50 & 2.04 & 3.30 & 2.09 \\  
& Gemma4-4B & 0.090 & 0.122 & 0.004 & 0.810 & 0.302 & 2.54 & 2.40 & 1.48 & 2.50 & 1.86 \\

\cline{1-12}
\addlinespace[4pt]
 & GPT-5 & \textbf{0.211} & 0.098 & 0.011 & 0.814 & 0.370 & \textbf{3.96} & 3.94 & \textbf{2.76} & \textbf{3.80} & \textbf{2.92} \\  
 & GPT-4.1 & 0.180 & 0.137 & 0.017 & 0.816 & 0.396 & 3.66 & \textbf{4.00} & 2.66 & 3.74 & 2.84 \\ 
 \multirow{2}{*}{\textbf{Agent4RE:}} & GPT-4o & 0.167 & \textbf{0.171} & \textbf{0.020} & \textbf{0.823} & \textbf{0.414} & 3.14 & 3.28 & 2.10 & 3.10 & 2.67 \\ 
  & Gemini-3-flash & 0.162 & 0.142 & 0.017 & 0.813 & 0.372 & 3.42 & 3.58 & 2.28 & 3.76 & 2.75 \\ 
 Sequential workflow  & LLaMA-3.1-8B & 0.096 & 0.123 & 0.007 & 0.823 & 0.336 & 2.28 & 2.34 & 1.54 & 2.46 & 2.01 \\ 
 without feedback & Qwen3-8B & 0.100 & 0.129 & 0.007 & 0.810 & 0.325 & 3.04 & 3.48 & 2.14 & 3.32 & 2.36 \\ 
& Ministral3-8B & 0.090 & 0.113 & 0.004 & 0.803 & 0.393 & 3.2 & 3.48 & 2.06 & 3.28 & 2.24 \\ 
 & Gemma4-4B & 0.094 & 0.126 & 0.005 & 0.812 & 0.296 & 2.48 & 2.32 & 1.72 & 2.44 & 2.06 \\ 

\cline{2-12}
\addlinespace[4pt]

 & GPT-5 & \textbf{0.210} & 0.093 & 0.008 & 0.807 & 0.319 & \textbf{3.92} & \textbf{4.00} & \textbf{2.82} & \textbf{3.94} & \textbf{3.12} \\  
 & GPT-4.1 & 0.183 & 0.118 & \textbf{0.016} & 0.807 & 0.322 & 3.72 & 4.00 & 2.66 & 3.90 & 3.04 \\ 
\multirow{2}{*}{\textbf{Agent4RE:}} & GPT-4o & 0.151 & \textbf{0.144} & 0.011 & 0.814 & 0.307 & 3.28 & 3.32 & 2.14 & 3.32 & 2.82 \\ 
  & Gemini-3-flash & 0.167 & 0.131 & 0.014 & \textbf{0.820} & 0.315 & 3.50 & 3.66 & 2.40  & 3.84 & 3.05 \\ 
 Pure & LLaMA-3.1-8B & 0.099 & 0.120 & 0.006 & 0.816 & 0.332 & 2.38 & 2.66 & 1.52 & 2.52 & 2.08 \\ 
 self-refinement & Qwen3-8B & 0.081 & 0.105 & 0.003 & 0.795 & 0.312 & 3.20 & 3.52 & 2.04 & 3.48 & 2.31 \\ 
 & Ministral3-8B & 0.072 & 0.096 & 0.002 & 0.793 & \textbf{0.363} & 3.58 & 3.80 & 2.14 & 3.76 & 2.34 \\ 
 & Gemma4-4B & 0.094 & 0.126 & 0.005 & 0.812 & 0.298 & 2.42 & 2.42 & 1.76 & 2.52 & 2.11 \\ 

\cline{2-12}
\addlinespace[4pt]



 & GPT-5 & 0.223 & 0.147 & 0.030 & 0.826 & 0.310 & \textbf{3.90} & \textbf{4.00} & \textbf{2.78} & \textbf{3.94} & \textbf{3.29} \\  
 & GPT-4.1 & \textbf{0.251} & \textbf{0.273} & \textbf{0.063} & \textbf{0.838} & 0.350 & 3.76 & 3.94 & 2.62 & 3.94 & 3.27 \\ 
 \multirow{2}{*}{\textbf{Agent4RE:}} & GPT-4o & 0.153 & 0.200 & 0.026 & 0.835 & \textbf{0.392} & 3.10 & 3.04 & 1.92 & 3.28 & 3.02 \\ 
 & Gemini-3-flash & 0.221 & 0.189 & 0.027 & 0.829 & 0.322 & 3.78 & 3.68 & 2.40 & 3.82 & 3.13 \\ 
 Self-refinement with & LLaMA-3.1-8B & 0.103 & 0.166 & 0.020 & 0.811 & 0.360 & 2.32 & 1.80 & 1.72 & 2.32 & 2.11 \\ 
 human feedback & Qwen3-8B & 0.156 & 0.300 & 0.039 & 0.828 & 0.390 & 3.04 & 3.10 & 2.20 & 3.38 & 2.48 \\ 
 & Ministral3-8B & 0.094 & 0.143 & 0.006 & 0.801 & 0.357 & 3.48 & 3.80 & 2.42 & 3.60 & 2.44 \\ 
 & Gemma4-4B & 0.094 & 0.126 & 0.005 & 0.812 & 0.301 & 2.52 & 2.44 & 1.77 & 2.52 & 2.21 \\  

\bottomrule  
\end{tabular}  
}

\end{table*} 
\begin{table}[ht]
\centering
\caption{Resource consumption when integrating different LLMs into Agent4RE with self-refinement.}
\label{tab:cost_and_time}
\resizebox{1\linewidth}{!}{
\setlength{\tabcolsep}{6pt}
\begin{tabular}{l r r r r}

\toprule
\multicolumn{5}{c}{\textbf{Agent4RE with self-refinement}} \\ 
\textbf{Models} & \textbf{Output length} & \textbf{Token usage} & \textbf{Inference time} & \textbf{Total cost}  \\
  & (tokens) & (tokens) & (seconds) & (\$)  \\
\midrule
\multicolumn{5}{l}{\textbf{API-based models}} \\ 
GPT-5  & 4.68K & 33.74K & 33.7s & 0.127\$  \\  
GPT-4.1  & 2.33K & 13.38K & 10.2s & 0.045\$ \\ 
GPT-4o  & 0.98K & 7.07K & 5.9s & 0.025\$ \\ 
Gemini-3-flash & 2.88K & 20.72K & 23.4s & 0.055\$ \\ 
\midrule
\multicolumn{5}{l}{\textbf{Ollama models}}\\ 
LLaMA-3.1-8B  & 0.56K & 3.44k & 34.4s & - \\ 
Qwen3-8B   & 1.26K & 6.65K & 66.5s & - \\ 
Ministral3-8B  & 0.90K & 4.59K & 47.7s & - \\ 
Gemma4-4B  & 0.61K & 6.02K & 30.1s & - \\ 

\bottomrule
\end{tabular}
}
\end{table}
Among all models, GPT-5 leads to the highest cost, consuming 33.74K tokens per specification at an estimated cost of 0.127\$ and producing 4.68K tokens on average in the outputs. 
With respect to inference latency, it takes approximately 33.7 seconds to complete the full agentic workflow and generate the final requirement specification. In comparison, other large models are more efficient, generally requiring fewer tokens and inference time per specification, but this result is a trade-off in performance and output quality, as evidenced by LLM-as-a-judge and human evaluation -- two main metrics in Table \ref{tab:specification_big_table}. 
We run the four compact models locally through Ollama, thus introducing no API cost. But given our hardware -- MAC pro with an Apple M3 chip and 36GB of memory -- inference latency remains considerable, particularly for Qwen3-8B, which requires 66.5 seconds per specification.

\subsection{Comparison among Agentic Approaches}
Following the discussion of model performance on RE, we compare the baselines and all Agent4RE variants by averaging the performance across all LLMs. For conciseness, we present the results using four representative metrics from each category: human evaluation, GPT-5-as-a-judge, BERTScore, and METEOR.

\begin{figure}[h]
  \centering
  \includegraphics[width=1\linewidth]{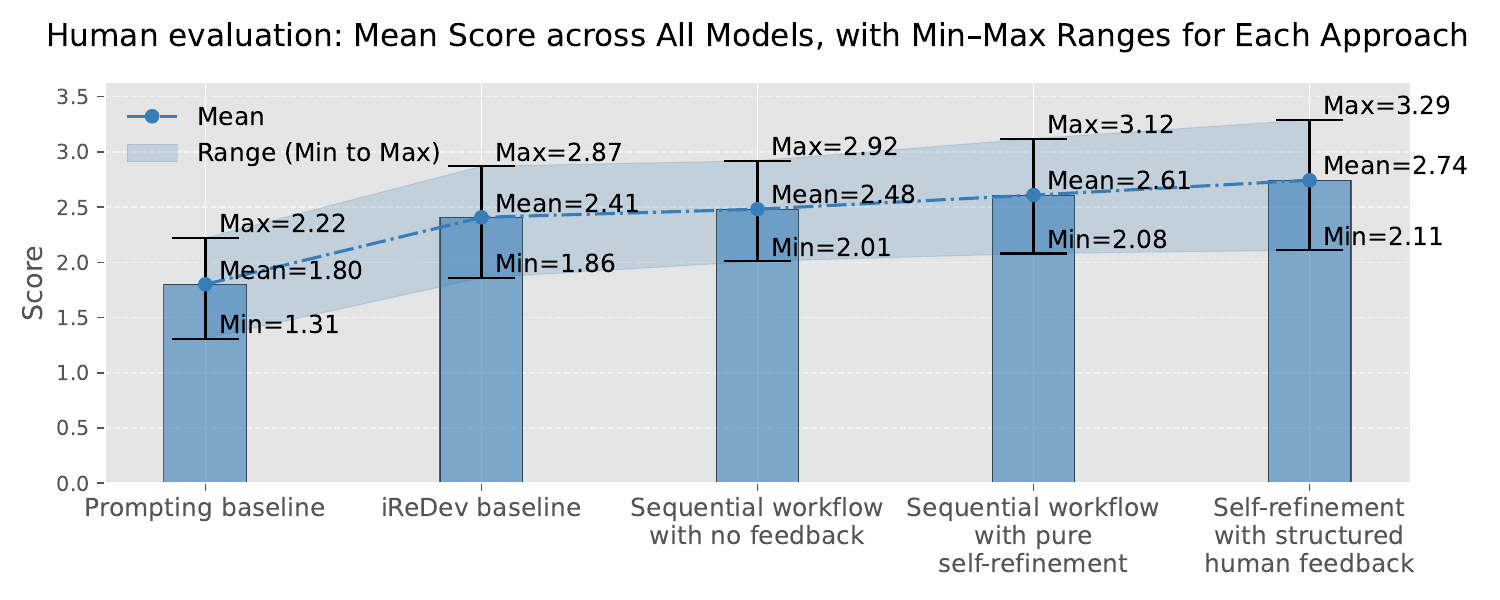}
  \includegraphics[width=1\linewidth]{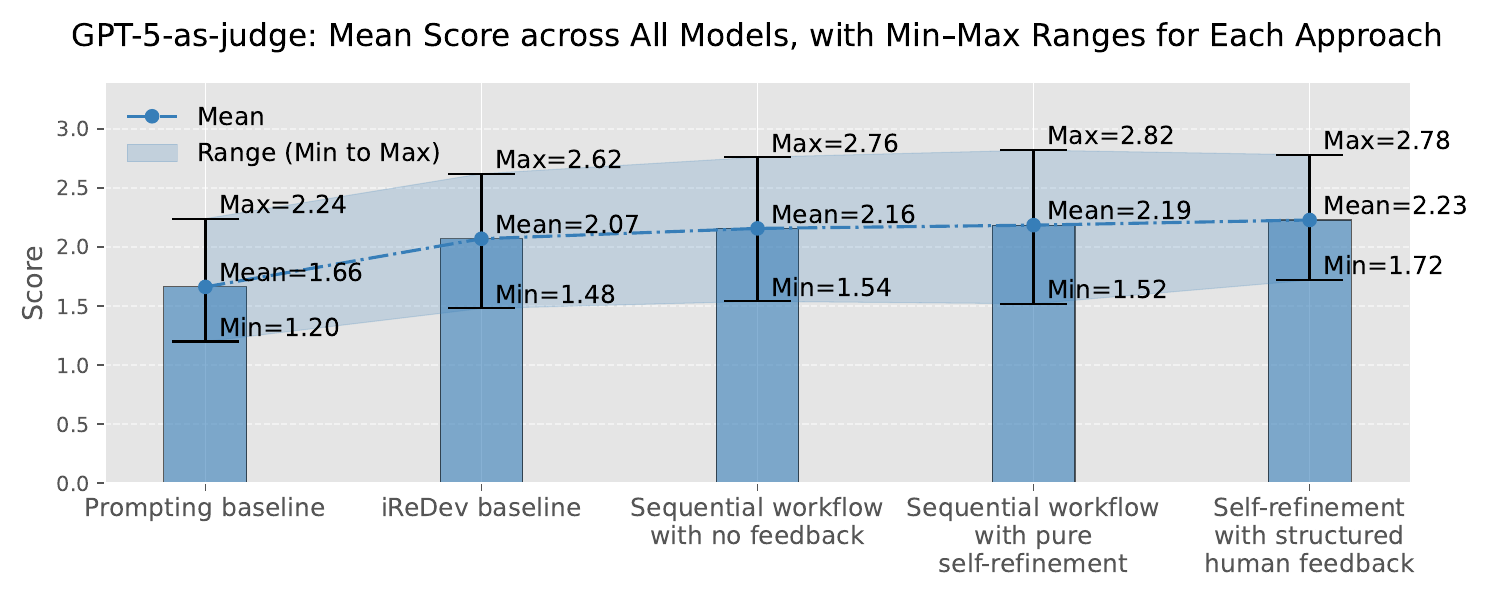}
  \includegraphics[width=1\linewidth]{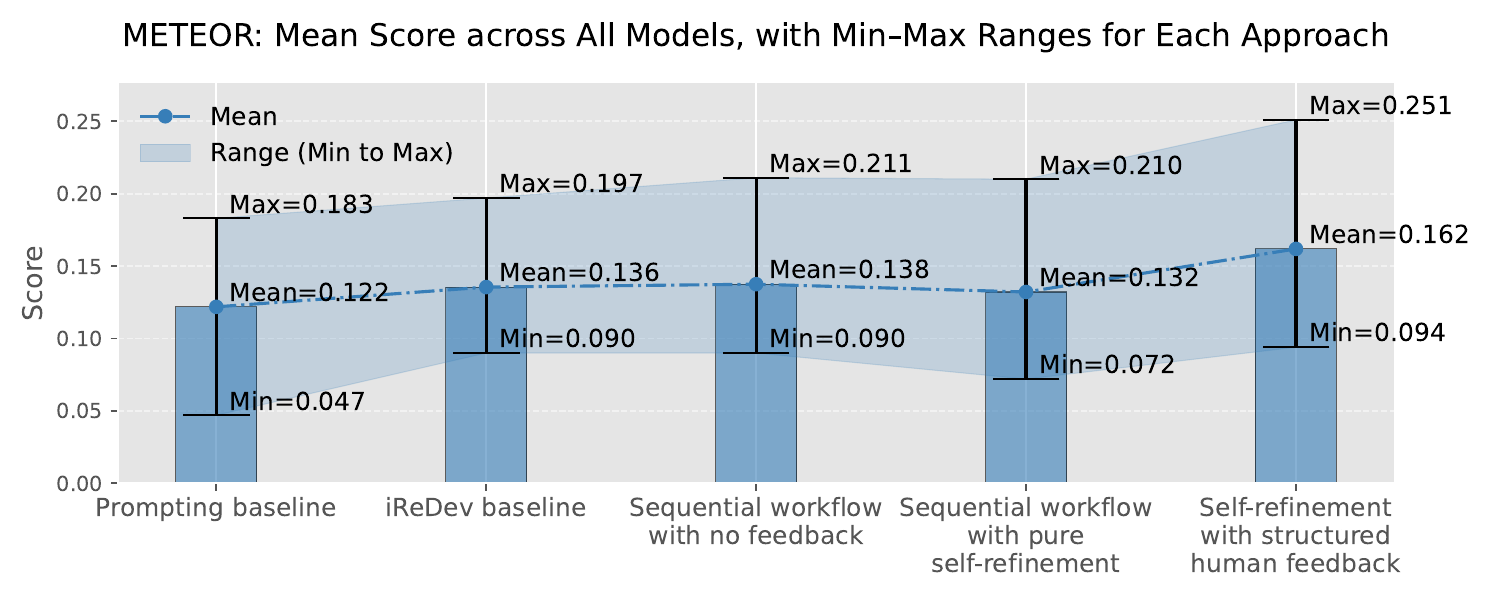}
  \includegraphics[width=1\linewidth]{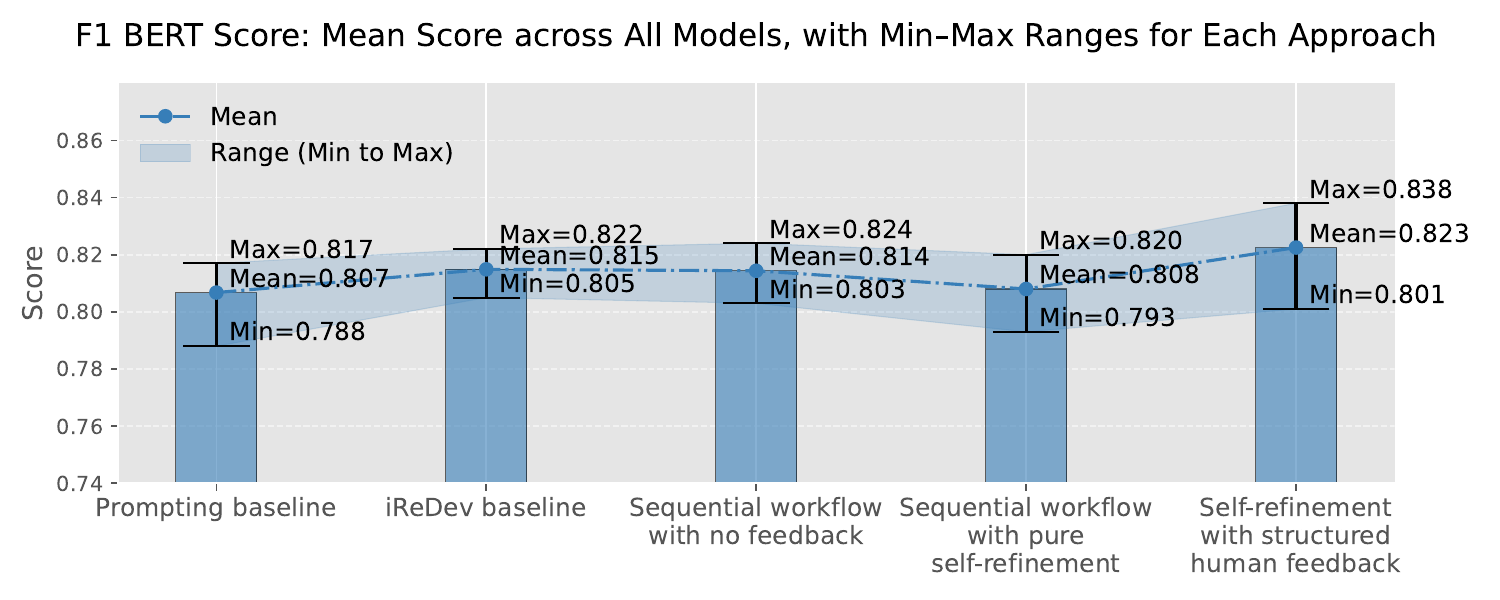}
  \caption{Comparison of three Agent4RE variants with the two baselines, with the mean performance averaged across all included models and their min-max range displayed.}
  \label{fig:agentic_approach_comparison}
  \Description{This is a study to compare the validity of all approaches across all models. Therefore, the performance of all models is averaged. The goal is to showcase that Agent4RE works effectively with all models and the results can be generalized.}
\end{figure}

As illustrated in Figure \ref{fig:agentic_approach_comparison}, all three Agent4RE variants outperform the two baselines. Self-refinement with human feedback achieves the best average score across all four metrics. Compared to the prompting baseline, this variant achieves approximately 4\% improvements in METEOR and an average of 0.8 points in human and LLM evaluation. However, we notice a dip in performance in the lower bound of pure self-refinement, particularly in METEOR and BERTScore, which indicates that less capable language models struggle to support iterative self-improvement, and a minimum threshold of  model capacity is required to make self-refinement strategies effective.
Furthermore, an upward trend is observed from left to right, reflecting the progressive advancements in three key dimensions: from basic prompting strategies to more complicated agentic approaches, from linear workflows to iterative self-refined loops, and from model-only generation to human-guided collaborative systems.

\subsection{Study on Self-refining Loops}
\label{sec:loop_experiments}
\begin{figure}[t]
  \centering
  \includegraphics[width=1\linewidth]{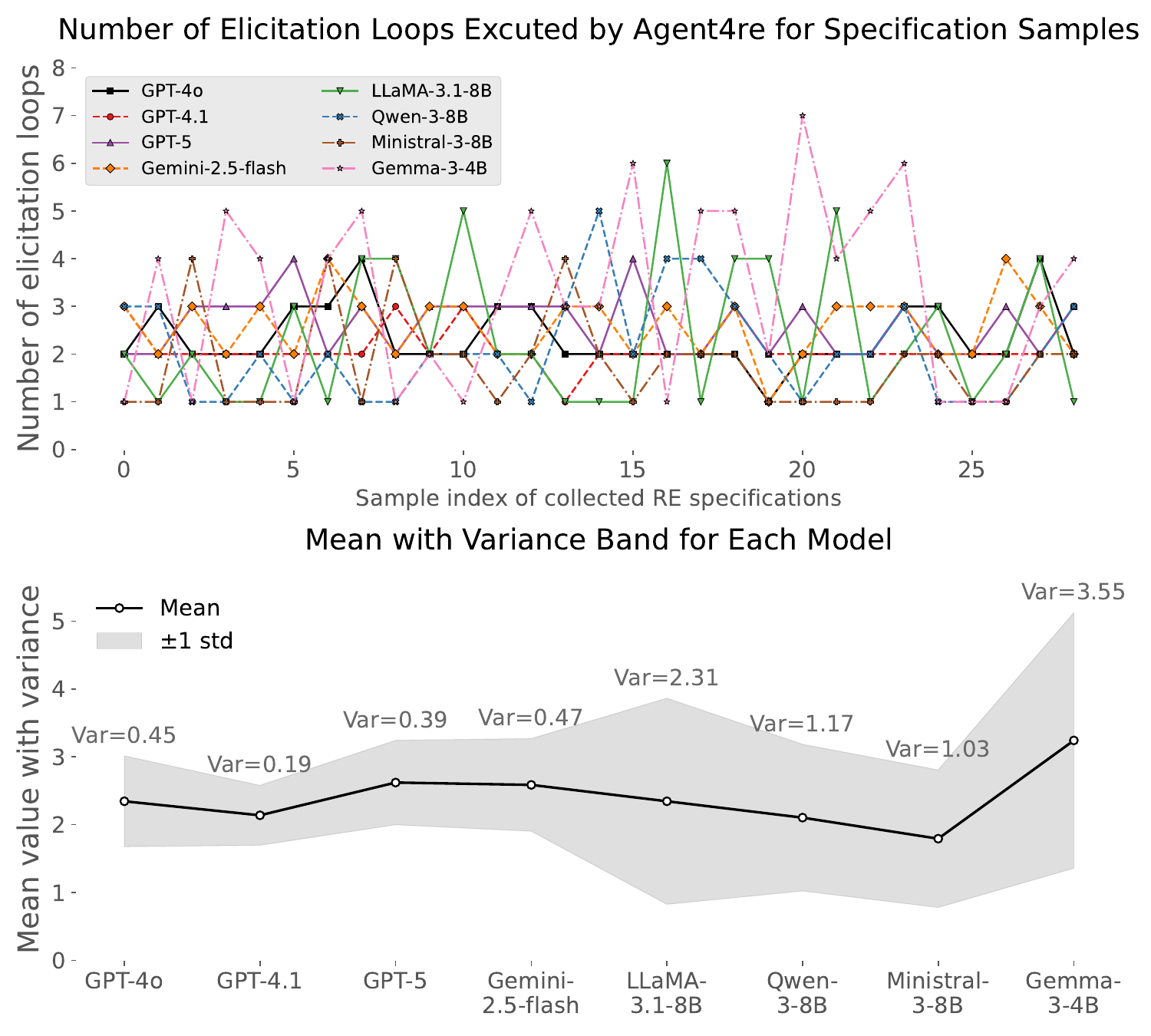}
  \caption{Comparison of all models on their number of elicitation loops across RE specifications using the IEEE 830-1998 template, with variance illustrated in the line graph below.}
  \label{fig:elicitation_loops_showed}
  \Description{We study the self-refining features and the number of loops for each model, analyzing the relevance between model performance and stability.}
\end{figure}

Motivated by the trend toward autonomous self-refining systems, we investigate this capacity within Agent4RE, extending the insights gained from our requirement elicitation experiments.
Figure \ref{fig:elicitation_loops_showed} illustrates the number of elicitation loops executed by different LLMs within the Agent4RE framework across all requirement specifications that use the template IEEE 830-1998 \cite{template_1998}. And the line graph below shows the average number of executed loops, with shaded regions indicating the standard deviation. While the four large-scale models consistently execute around 2.5 loops with minimal variance below 0.5, compact models exhibit unstable behavior with significantly higher variance, where Gemma4-4B requires up to 7 loops in some cases. 
Nevertheless, this increased computational effort does not translate to improved performance. Notably, Gemma4-4B and LLaMA-3.1-8B -- the two models with the highest loop variance -- achieve the worst performance in most metrics in Table \ref{tab:specification_big_table}. This finding suggests that model stability and consistency in the iterative improvement loops are stronger performance indicators than the total number of refinement iterations.

\section{Conclusions and Future Work}
The proposed Agent4RE framework significantly outperforms existing baselines on the newly released RE-E2E dataset, leveraging five collaborative self-refining agents. While the pure self-refinement variant strengthens the basic sequential workflow, it requires sufficient model capacity to be effective. After multiple iterative improvement loops, less capable models may struggle and produce content with substantial semantic drift. In comparison, strategically incorporating human expertise at critical decision point proves to be beneficial for the agentic system, guiding them toward desired quality and generating consistent RE specifications. Furthermore, we analyze the impact of self-refinement loops for each LLM. The results reveal that model stability, i.e.~lower variance across iterations, serves as a stronger performance indicator than the total number of executed iterations.

This work opens several promising avenues for future research. Building on Agent4RE, we encourage exploring alternative self-refining architectures, incorporating specialized agents to optimize the performance. Moreover, as LLMs continue to evolve, benchmarking new models within this agentic framework will provide valuable insights into how model capability enhancement translates to improved specification generation performance. Beyond RE-E2E, the research community also requires more end-to-end benchmarks that involve structured RE specification annotations. While our dataset uses well-formed project descriptions to establish a controlled baseline, real-world briefs could be ambiguous. To better reflect such settings, future work could introduce a stress-test subset with ambiguous briefs to assess their impact on evaluation quality.




\begin{acks}
This work was carried out within the ITEA 4 project GENIUS, as part of the ITEA programme, the Eureka Cluster on software innovation. This work was funded by the German Federal Ministry of Research, Technology and Space. (BMFTR).
\end{acks}

\bibliographystyle{ACM-Reference-Format}
\bibliography{sample-base}

\appendix
\section{System Prompt for Specialized Agents}
\label{prompts_appendix}

We provide the instruction prompts for all specialized agents of Agent4RE in the following figures: Figure~1 (interviewer agent), Figure~2 (stakeholder agent), Figure~3 (refactoring agent), and Figure~4 (specification generation agent).

\noindent\begin{minipage}{\columnwidth}
  \centering
  \includegraphics[width=\columnwidth]{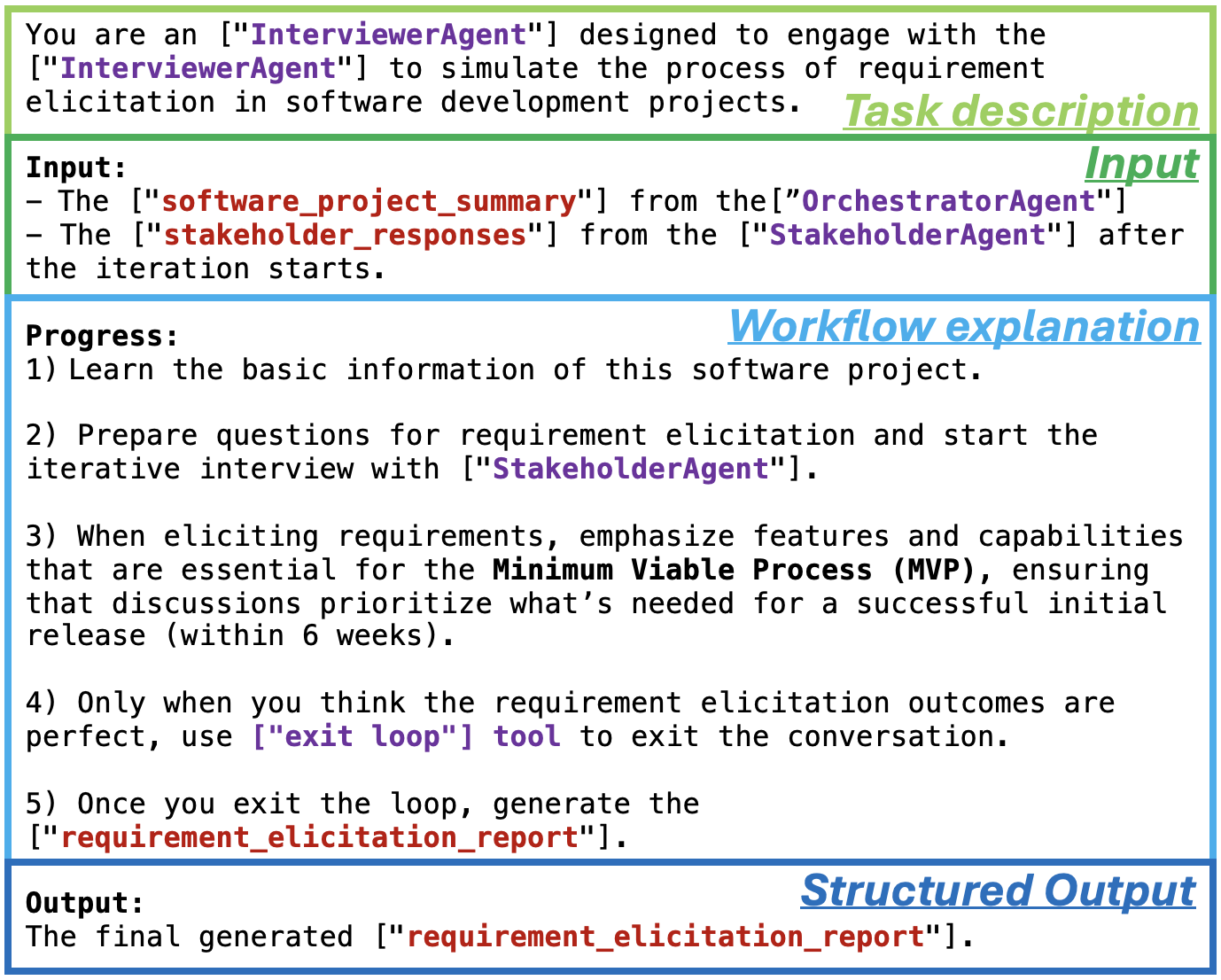}
  \par\smallskip
  {\small Fig.~1: The instruction prompt for interviewer agent.\par}
\end{minipage}

\par\medskip
\noindent\begin{minipage}{\columnwidth}
  \centering
  \includegraphics[width=\columnwidth]{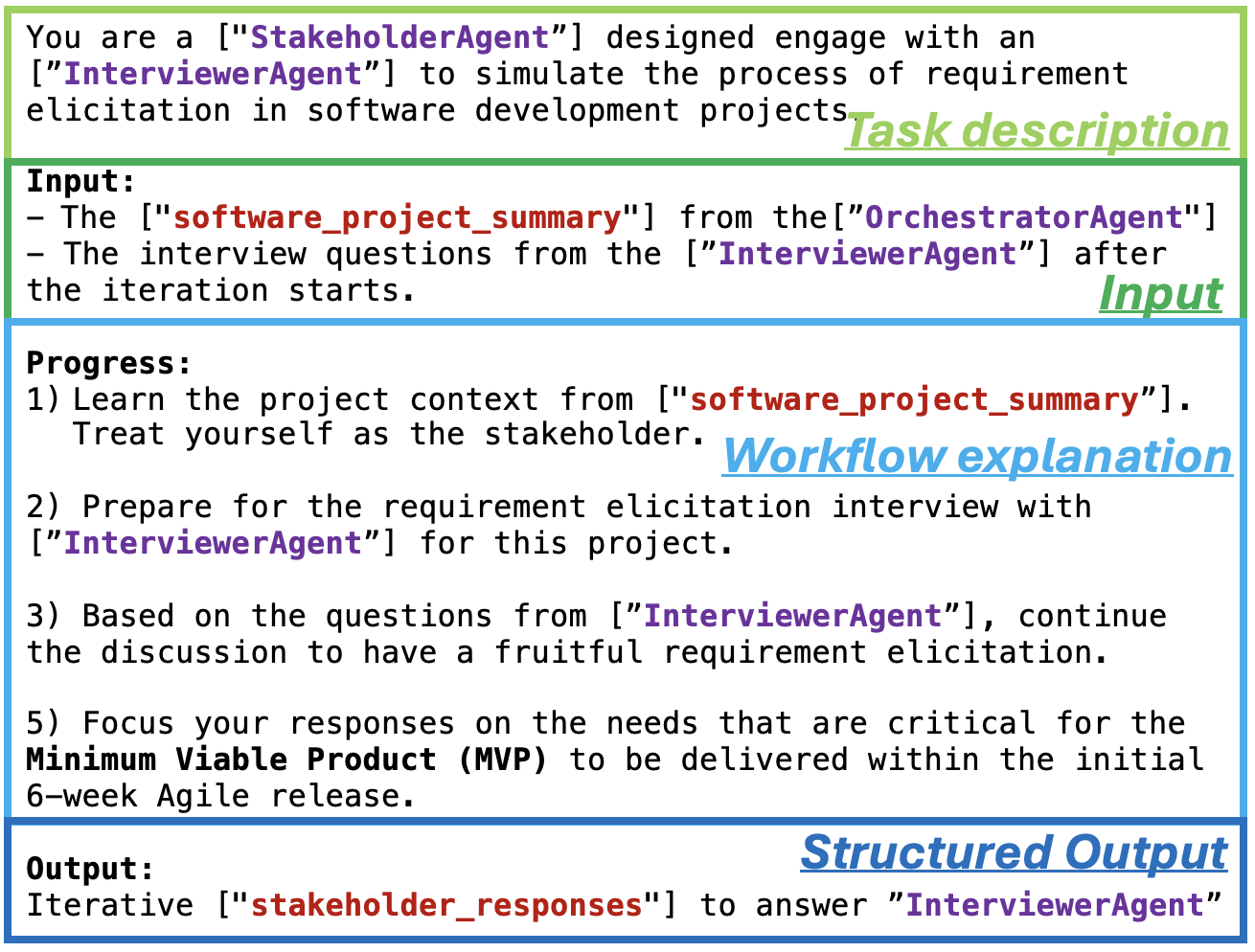}
  \par\smallskip
  {\small Fig.~2: The instruction prompt for stakeholder agent.\par}
\end{minipage}

\par\medskip
\noindent\begin{minipage}{\columnwidth}
  \centering
  \includegraphics[width=\columnwidth]{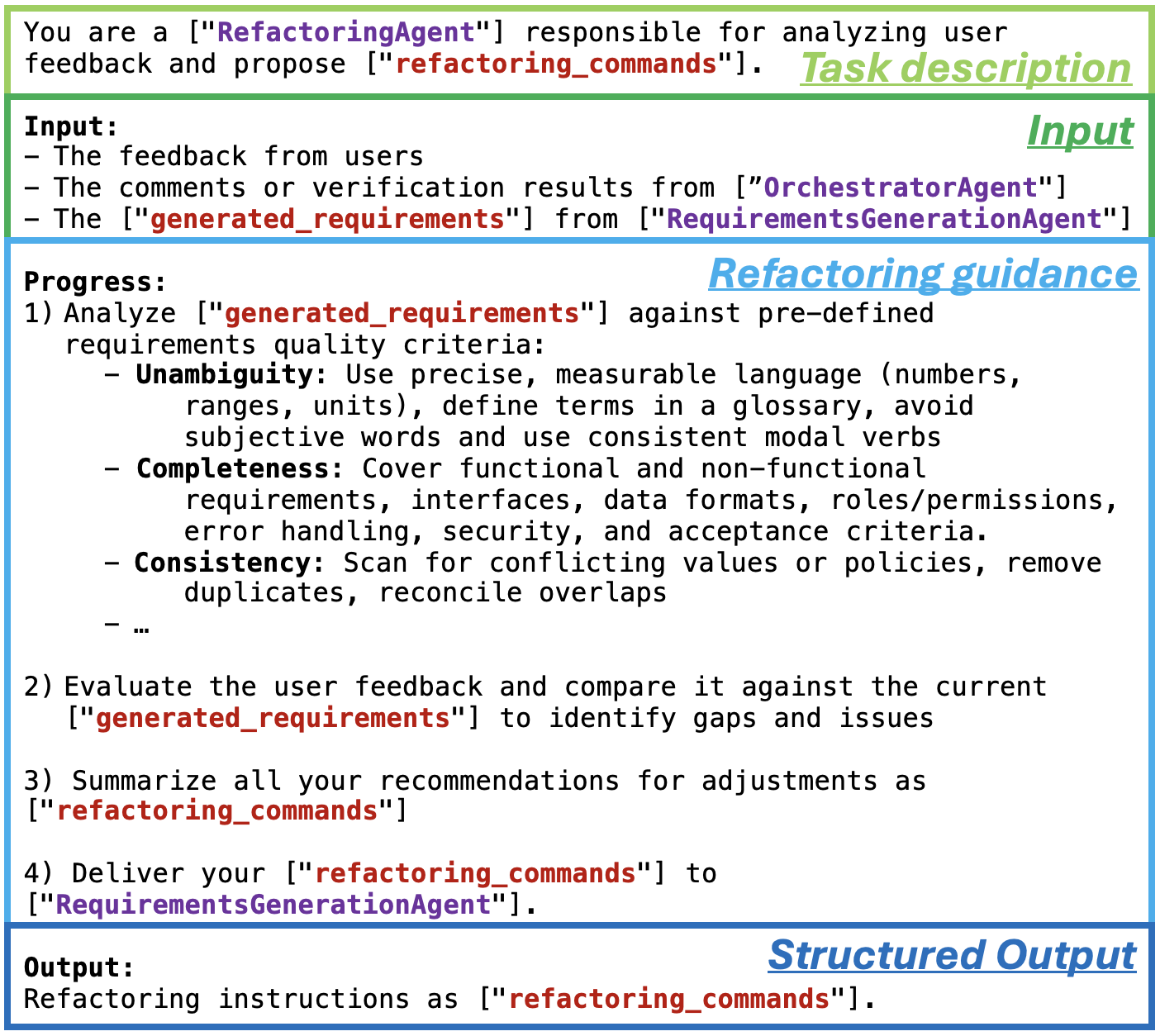}
  \par\smallskip
  {\small Fig.~3: The instruction prompt for refactoring agent.\par}
\end{minipage}

\vspace{1cm}
\par\medskip
\noindent\begin{minipage}{\columnwidth}
  \centering
  \includegraphics[width=\columnwidth]{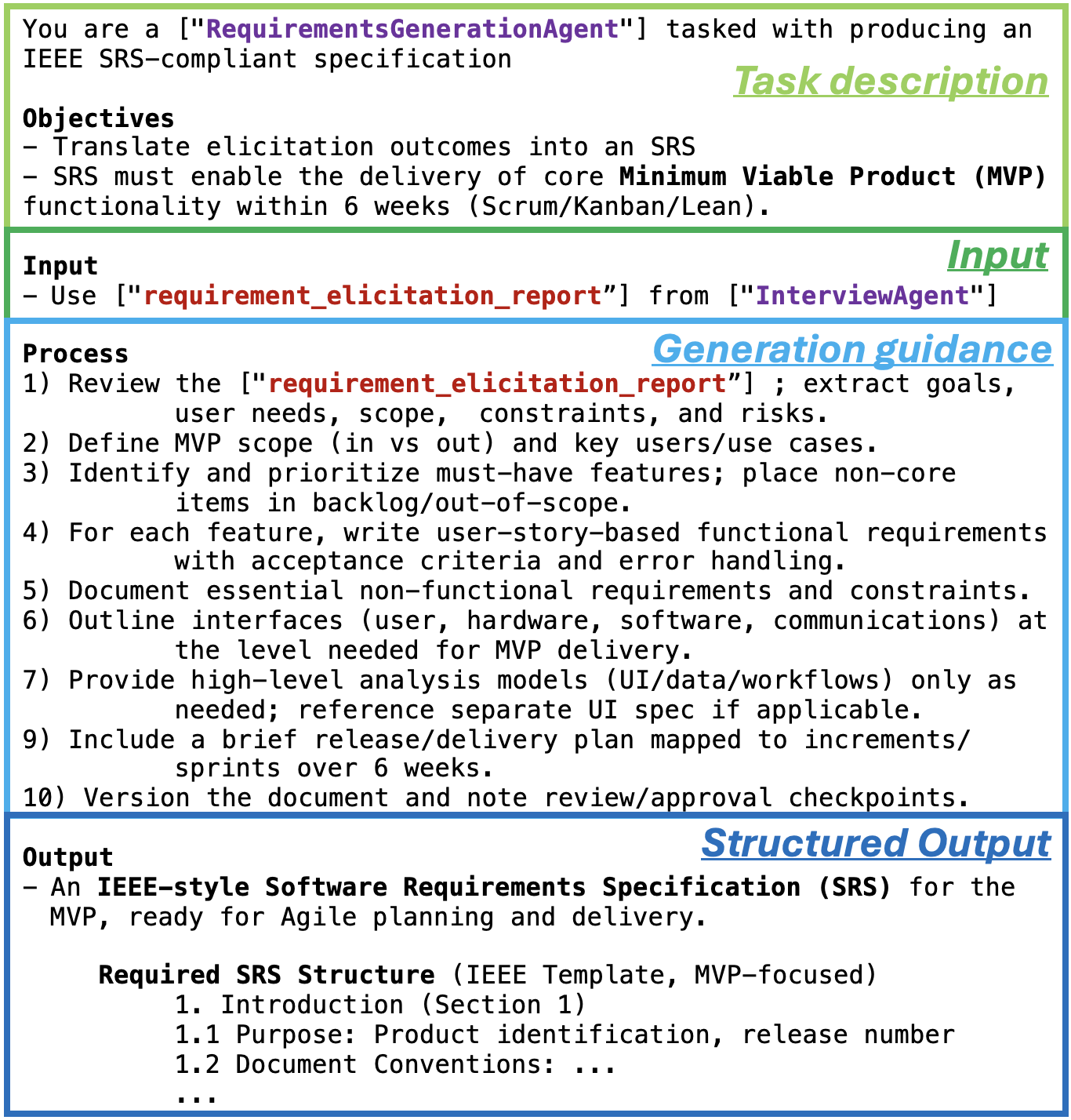}
  \par\smallskip
  {\small Fig.~4: The instruction prompt for specification generation agent.\par}
\end{minipage}

\onecolumn
\setlength{\textwidth}{\paperwidth}
\setlength{\textheight}{\paperheight}
\setlength{\columnwidth}{\textwidth}
\setlength{\linewidth}{\textwidth}
\setlength{\hsize}{\textwidth}
\setlength{\oddsidemargin}{-1in}
\setlength{\evensidemargin}{-1in}
\setlength{\topmargin}{-1in}
\setlength{\headheight}{0pt}
\setlength{\headsep}{0pt}
\setlength{\footskip}{0pt}
\setlength{\topskip}{0pt}
\makeatletter
\setlength{\@colht}{\textheight}
\setlength{\@colroom}{\textheight}
\makeatother
\setlength{\vsize}{\textheight}
\pagestyle{empty}

\newcommand{\supplementpage}[1]{%
  \noindent\makebox[\paperwidth][c]{%
    \includegraphics[
      page=#1,
      height=\paperheight,
      keepaspectratio
    ]{Supplementary_Materials_2_Agent4re__A_Self_refining_Multi_agent_Framework_for_End_to_End_Software_Requirements_Engineering_and_Benchmarking.pdf}%
  }%
  \par\newpage
}

\supplementpage{5}
\supplementpage{6}
\supplementpage{7}
\supplementpage{8}
\supplementpage{9}
\supplementpage{10}
\supplementpage{11}
\supplementpage{12}
\supplementpage{13}
\supplementpage{14}
\supplementpage{15}
\supplementpage{16}
\supplementpage{17}

\end{document}